# Engineering the optoelectronic properties of MoS$_2$ photodetectors through reversible noncovalent functionalization

Aday J. Molina-Mendoza,[a] Luis Vaquero-Garzon,[b] Sofia Leret,[b] Leire de Juan-Fernández,[b] Emilio M. Pérez,[b,]* and Andres Castellanos-Gomez.[b]*

[a] *Departamento de Física de la Materia Condensada, Universidad Autónoma de Madrid, Campus de Cantoblanco, E-28049, Madrid, Spain.*

[b] *IMDEA Nanociencia, C/Faraday 9, Campus de Cantoblanco, E-28049 Madrid, Spain.*

E-mail: andres.castellanos@imdea.org, emilio.perez@imdea.org

**Abstract**

We present an easy drop-casting based functionalization of MoS$_2$-based photodetectors that results in an enhancement of the photoresponse of about four orders of magnitude, reaching responsivities up to 100 A·W$^{-1}$. The functionalization is technologically trivial, air-stable, fully reversible and reproducible, and opens the door to the combination of 2D-materials with molecular dyes for the development of high performance photodetectors.

Among the novel two-dimensional (2D) materials,[1-8] transition metal dichalcogenides (TMDCs)[9-12] show particularly promising electronic and optoelectronic properties.[13] In particular, their intrinsic bandgap within the visible part of the spectrum, makes them highly interesting materials for optoelectronic applications.[14] In fact, the presence of a bandgap has allowed for the construction of a wealth of prototype electronic devices based on TMDCs.[15-24] In the last years, there has been a significant effort to modulate the optical properties of TMDCs in order to optimize the performance of the corresponding devices. Most of the strategies investigated so far rely on physical methods, such as strain-engineering,[25, 26] field-effect doping,[12, 27] or artificial stacking of different 2D materials.[28, 29] In comparison, the chemical modification of TMDCs is still rather underexplored, despite the appealing combination of low-cost and high degree of control offered by synthetic chemistry.
Examples of doping of TMDCs through surface charge-transfer using metal atoms,[30] gases,[31] and a few organic molecules has already been demonstrated.[32, 33] Responsivities of just a few A·W$^{-1}$ have been reported for MoS$_2$ photodetectors functionalized with a rhodamine dye,[34] a rather modest value for MoS$_2$-based photodetectors. Among the readily available organic dyes, perylenediimides (PDIs) and porphyrins show remarkable optical properties, including large molar absorptivity —ca. $10^5$ M$^{-1}$ cm$^{-1}$ for PDIs and $10^6$ M$^{-1}$ cm$^{-1}$ for porphyrins—, and outstanding photostability under ambient conditions.
These intrinsic properties have made them two of the most popular families of organic dyes, particularly in the frame of photovoltaics.[35-44] However, their use for the modulation of the optoelectronic properties of TMDC-based devices has not been yet described. Considering this, we decided to investigate the effects of the noncovalent functionalization of MoS$_2$ photodetectors with the soluble PDI and tetraphenyl porphyrin (TPP) depicted in Chart 1. Here, we describe that the supramolecular functionalization of mechanically exfoliated MoS$_2$-based photodetectors with PDI and TPP leads to an enhancement of photocurrent generation of over four orders of magnitude, making our functionalized devices highly sensitive (responsivities up to 100 A·W$^{-1}$). The process is technologically trivial, air-stable, reproducible and fully reversible.

MoS$_2$ phototransistors are fabricated by deterministic transfer of mechanically exfoliated flakes onto pre-patterned drain and source electrodes fabricated by shadow mask evaporation of Ti/Au (5 nm / 50 nm) onto a SiO$_2$ (285 nm in thickness) substrate, thermally grown on a highly p-doped silicon chip.[45]

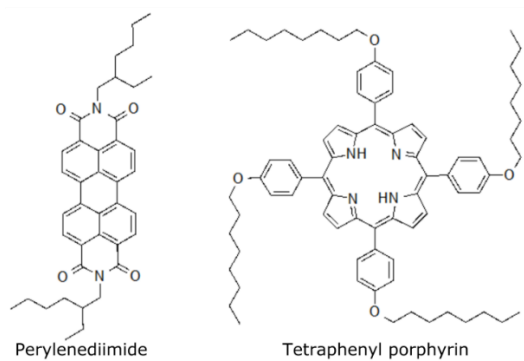

Chart 1. Chemical structure of the perylenediimide and tetraphenyl porphyrin investigated.

The heavily doped silicon is employed as a back gate electrode to tune the density of charge carriers by electric field- effect. Figure 1a-1 shows an artistic representation of a MoS$_2$-based field-effect transistor (FET) fabricated by transferring a MoS$_2$ flake bridging the drain and source electrodes. Figure 1b-1 shows an actual optical microscopy image of a fabricated device composed of a multilayer (~10 layers) MoS$_2$ flake. Right after the fabrication, we perform the FET characterization of the device in air and in dark conditions by sweeping the drain-source voltage and measuring the current passing through the MoS$_2$ flake. Figure 1c-1 shows a set of current vs. drain-source voltage curves (*IV*s hereafter) acquired at different applied gate voltages. The *IV*s show a typical n-type FET behavior, i.e. the current increases with increasing back-gate voltage. The threshold voltage is ~ 0 V and the mobility ~ $2 \times 10^{-3}$ cm$^2 \cdot$V$^{-1} \cdot$s$^{-1}$, although this value is lower bound since we are measuring with two terminals and a correction, which usually yields higher mobility values (typically ranging from 3 to 100 cm$^2 \cdot$V$^{-1} \cdot$s$^{-1}$ [46-50]), should be taken into account.

Once the pristine device is electronically characterized, we functionalize the surface of the MoS$_2$ device by drop-casting a CH$_2$Cl$_2$ solution containing either PDI or TPP (Figure 1a-2, see the Supporting Information for details on the synthesis of the molecules). The drop is placed on the device, covering both the MoS$_2$ flake and the electrodes, and it is kept in air until the solvent evaporates, resulting in a uniform coverage of the device with crystallites (Figure 1b-2).

After deposition of the dyes, the electronic characterization of the device is performed again in air and dark conditions. Figure 1c-2 shows the *IV*s acquired for different back-gate voltages with the molecule-coated device. As shown, the current is increased with respect to the pristine device by almost three orders of magnitude due to the doping created by the molecules that donate electrons to the MoS₂ flake. Despite the typically electron-acceptor character of PDI, both molecules behave as electron donors when faced with MoS₂. Similar behavior has been observed even with positively charged (i.e. very strongly electron-accepting) organic molecules.[32] The result of this electron transfer process is an increase in the n-doping of the material, as can be depicted from the shift in the threshold voltage in the current-gate voltage traces (Figure S1). We find that the mobility of the device is also increased from $3\times10^{-3}$ cm²·V⁻¹·s⁻¹ in the pristine device up to ~ 1 cm²·V⁻¹·s⁻¹ both in the TPP and PDI functionalized devices (more field-effect characteristics of the device are listed in Table S1 of the Supporting Information). The current and mobility enhancement can be also due to the reduction of the Schottky barrier assisted by the molecular doping, as previously reported for MoS₂ FETs doped with Cl, an electron donor, particularly taking into account that we use a chlorinated solvent.[51] However, the difference in results obtained with each molecule, and most convincingly, the fact that we can return to the initial state by washing with $CH_2Cl_2$ (vide infra) unambiguously support a molecule-specific doping mechanism. It is important to note that in the coated devices the molecules do not contribute significantly to the total current passing through the device (Figure S2). Finally, the device is washed with $CH_2Cl_2$ in order to remove the molecules, as illustrated in Figure 1a-3. The device almost recovers the initial characteristics (Figure 1b-3), although the current is still slightly enhanced (Figure 1c-3), probably due to the presence of traces of molecules still adsorbed on the MoS₂ surface, or due to the solvent effect mentioned earlier.

We not only characterize the device in dark conditions, but also upon illumination. In Figure 2a we show a sketch of the optoelectronic measurement: the device is illuminated with a light beam guided through an optical fiber (the spot on the device has 200 µm in diameter, illuminating the whole device). The light is switched on and off while measuring the current. In Figure 2b and 2c we show the photocurrent ($I_{ph}$, the difference between the current upon illumination and in dark) as a function of time when the device is coated with PDI (Figure 2b, left) and with TPP (Figure 2c, left). The photocurrent is larger for the PDI coating than for the TPP coating due to the higher doping of the device by PDI and a better match of its absorption with the light source (see UV-vis spectra and Figure S3 in the Supporting Information). In Figure 2b and 2c, right, we show the same measurements after cleaning the device with $CH_2Cl_2$ in order to remove the molecular coating. The functionalized devices show a most remarkable enhancement of the photocurrent of $2\times10^4$ times for the TPP coating and $3\times10^4$ times for PDI. Note that scanning photocurrent (Figure S4) and wavelength dependent photocurrent measurements (Figure S5) show that the photocurrent is only generated when the MoS₂ is illuminated, and the MoS₂ absorption dominates over the molecules absorption (the photocurrent spectra after functionalization do not show strong spectral features due to the coating).

In order to study the reproducibility and reversibility of the functionalization/cleaning process, we repeat the characterization several times in the same device, not only coating with one molecule, but also changing the molecule. In Figure 3a we show a plot of the responsivity ($R$), a figure-of-merit for photodetectors that represents the input-output gain of the device for a given light power. The responsivity is defined as $R = I_{ph} / P_{eff}$, where $P_{eff}$ is the effective light power reaching the device and is calculated as $P_{eff} = P \cdot A_{dev}/A_{spot}$, with $P$ the light power reaching the substrate, $A_{dev}$ the area of the flake between the two electrodes and $A_{spot}$ the area of the spot on the device.

As it is shown in Figure 3a, the responsivity of the device is enhanced when it is functionalized with PDI ($\sim 10^2$ A·W$^{-1}$) three orders of magnitude with respect to the pristine device ($\sim 10^{-1}$ A·W$^{-1}$). When the device is cleaned, we recover a similar situation as in the pristine device with a responsivity of $\sim 10^{-2}$ A·W$^{-1}$. The process is then repeated twice using the same device, obtaining a similar value for the responsivity with a PDI coating and in the cleaned device.

When the device is functionalized with TPP, the responsivity oscillates between $10^2$ A·W$^{-1}$ and $10^1$ A·W$^{-1}$, and after cleaning the responsivity drops to $\sim 10^{-1}$ A·W$^{-1}$. The dramatic enhancement of the responsivity originates from the combination of photogating effect and a high photoconductive gain. The photogating effect is translated in the trapping of photogenerated charges in localized states within the bandgap of the material.[46, 47] In this case, these states are introduced at the surface of MoS₂ by the functionalization with PDIs or TPPs, making photogenerated holes to get trap in these localized states where they reside for long times, thus quenching the electron-hole recombination and, consequently, increasing the photoconductive gain. Together with the photogating mechanism, hybrid MoS₂/molecular dye systems are expected to present a high photoconductive gain because of the short carrier transist time ($\tau_{transist}$), due to the high mobility of MoS₂, and the long trapping lifetime in the dye molecule ($\tau_{lifetime}$): the gain is $\tau_{lifetime}/\tau_{transist}$.[21, 52, 53]

We also investigate the rise/fall time of the device with and without coating, defined as the time measured between 10% and 90% of the maximum photocurrent in the rising or falling edge. Here we observe that the rise/fall time increases from $\sim 1$ s for the pristine and cleaned devices to $\sim 10$ seconds for the PDI-coated device and to $\sim 300$ seconds for the TPP-coated device (Figure 3b). This indicates that the trapping lifetime of photogenerated holes due to the presence of the dye molecules is 1-2 orders of magnitude longer than those of intrinsic traps in pristine MoS₂ devices, in good agreement with a much higher photoconductive gain in the functionalized devices.

In summary, we have effectively functionalized MoS₂-based photodetectors with PDIs and TPPs by a simple drop-casting method, resulting in a dramatic enhancement of the photoresponse and responsivity (three orders of magnitude higher) to visible light. The process is fully reversible and reproducible, demonstrating that molecular dyes are an easy and interesting way of improving the performance of photodetectors based on two-dimensional materials.

## Acknowledgements

This work was supported by the BBVA Foundation through the fellowship "I Convocatoria de Ayudas Fundacion BBVA a Investigadores, Innovadores y Creadores Culturales", the MICINN (MAT2014-58399-JIN), the European Commission under the Graphene Flagship, contract CNECTICT-604391, the European Research Council (ERC-StG-307609, MINT) and the Spanish MINECO (Ramón y Cajal 2014 program, RYC-2014-01406 and CTQ2014-60541-P). A.J.M-M. acknowledges the financial support of MICINN (Spain) through the scholarship BES-2012-057346. L.J-F. is thankful to the MECD for an FPU scholarship (FPU13/03371).

## Figures

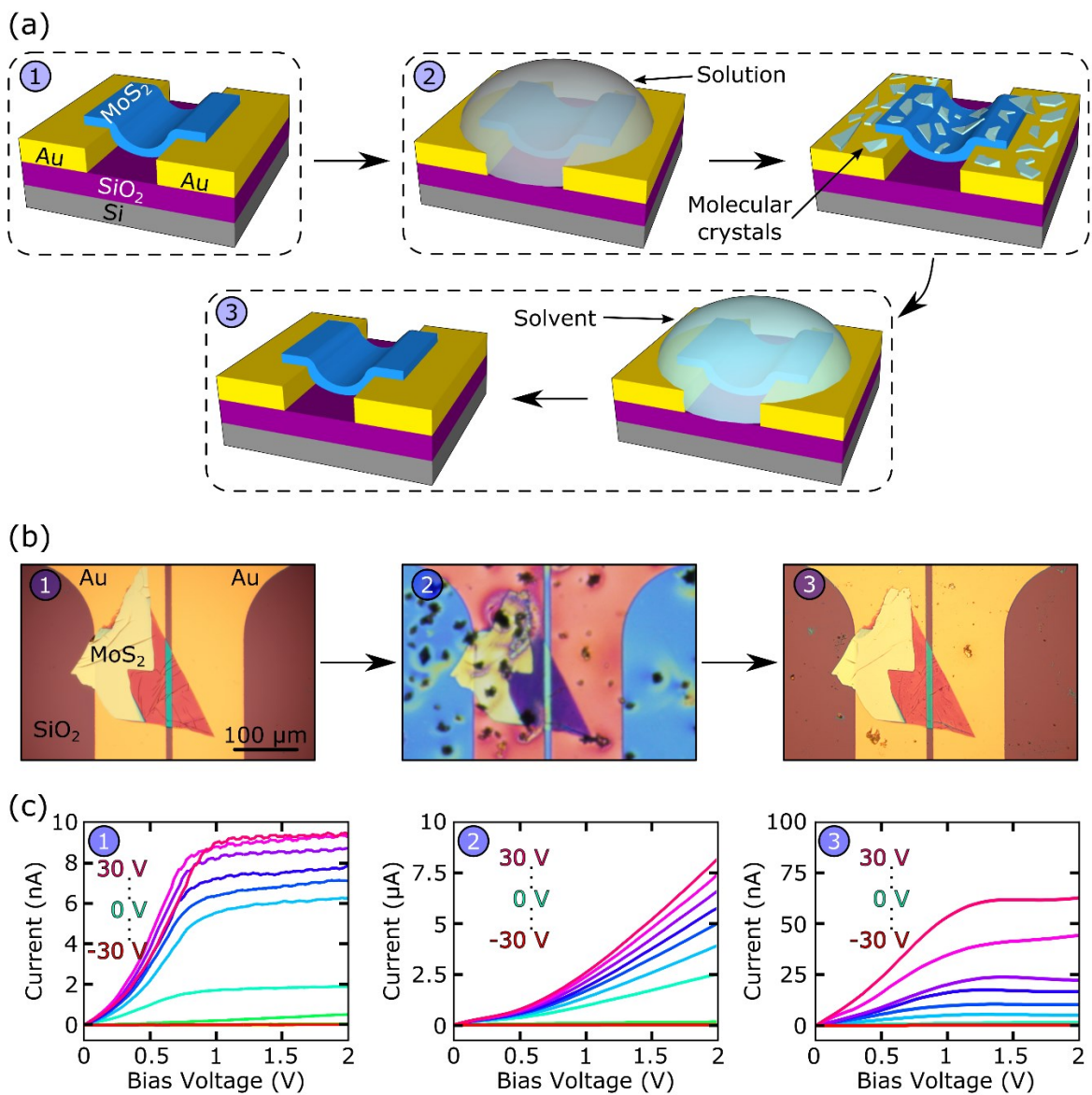

**Figure 1. (a)** Artistic representation of the fabrication of a MoS₂ phototransistor device subjected to one functionalization-cleaning cycle. (1) The pristine device is fabricated: a few-layer MoS₂ flake is placed bridging two Au electrodes by deterministic transfer. (2) Molecular crystals are deposited on the device by drop-casting. (3) The device is washed with solvent to remove the deposited molecules, resulting in a clean device with similar performance as the pristine one. **(b)** Optical microscopy images of the process described in (a): (1) Pristine device. (2) The same device covered with a layer of molecules (TPP in this case). (3) The same device after removing the molecules. **(c)**

*IV*s of the device shown in (b) for different back-gate voltages (from -30 V to 30 V) at the different steps of the process. Remarkably, after molecules deposition the current flowing through the device increases by three orders of magnitude.

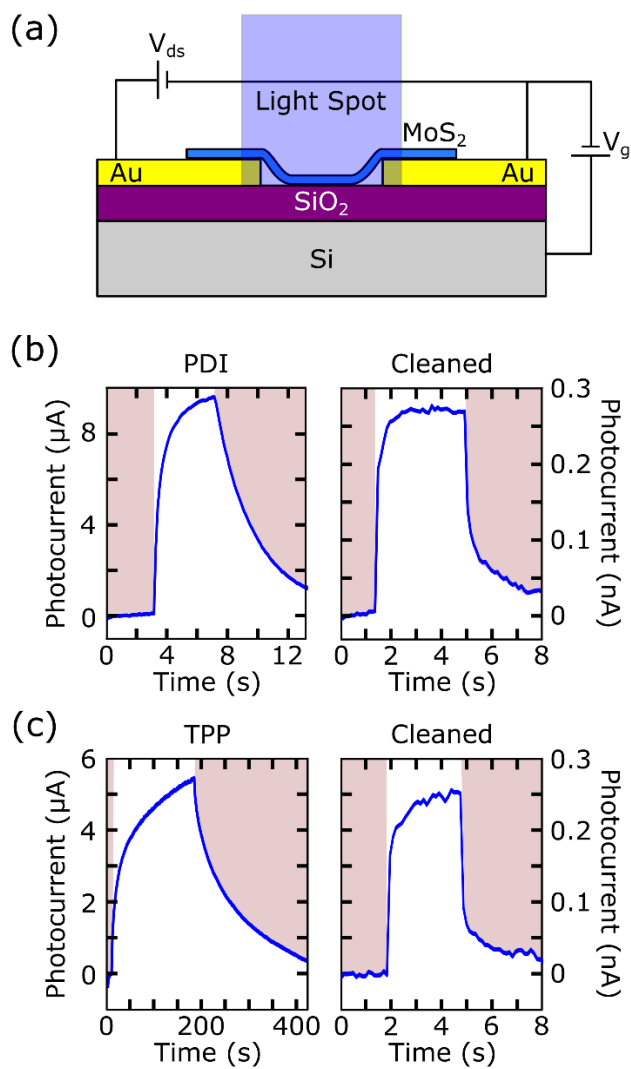

**Figure 2. (a)** Schematic drawing of the measurement method: a light spot (typically a LED beam guided by an optical fiber) illuminates the whole flake while the current passing through the MoS$_2$ flake is measured. The light is switched on and off to measure the photocurrent. **(b)** Photocurrent measured for the device shown in Figure 1b functionalized with PDI (left) and after cleaning the molecules (right) with light wavelength of 455 nm and power density of 72 mW/cm$^2$ ($V_{ds}$ = 2 V, $V_g$ = 30 V). The PDI functionalized device photoresponse is 3×10$^4$ times larger than that of the cleaned device. **(c)** Photocurrent measured for the device shown in Figure 1b covered by TPP (left) and after removing the molecules (right) with light wavelength of 455 nm and power density of 72 mW/cm$^2$ ($V_{ds}$ = 2 V, $V_g$ = 30 V). The functionalized device photoresponse is 2×10$^4$ times larger than after being cleaned.

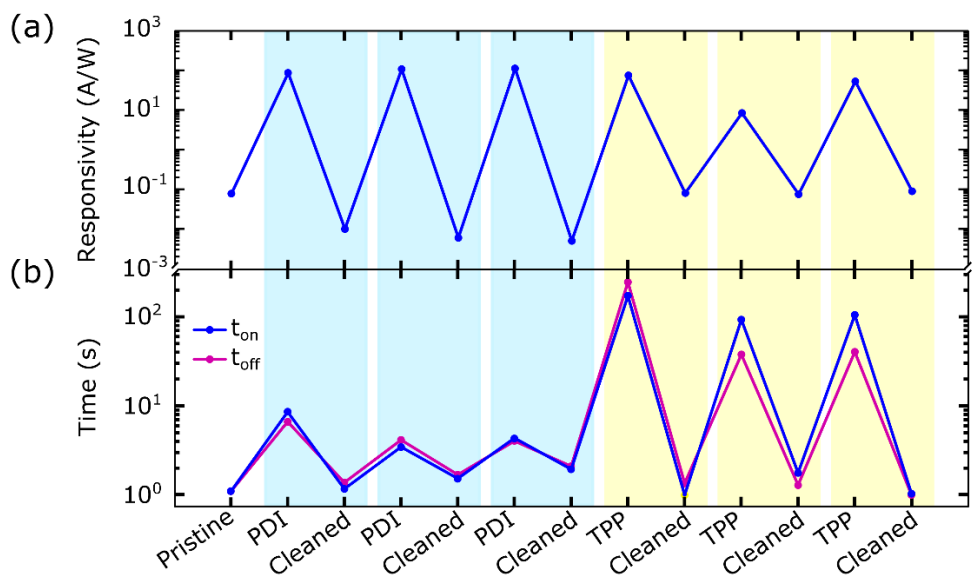

**Figure 3. (a)** Responsivity for the different cycles of chromophore functionalization and solvent cleaning. The higher responsivity values are achieved when the device is covered with PDI. The LED wavelength is 455 nm, and the power density is 43 mW/cm$^2$. **(b)** Time response for the different cycles. The fastest response is achieved with the pristine and cleaned devices, but the time response with PDI is just one order of magnitude larger than the pristine device.

## Supporting Information

## Engineering the optoelectronic properties of MoS$_2$ photodetectors through reversible noncovalent functionalization

Aday J. Molina-Mendoza,[a] Luis Vaquero-Garzon,[b] Sofia Leret,[b] Leire de Juan-Fernández,[b] Emilio M. Pérez,[b,*] and Andres Castellanos-Gomez.[b*]

[a] *Departamento de Física de la Materia Condensada, Universidad Autónoma de Madrid, Campus de Cantoblanco, E-28049, Madrid, Spain.*

[b] *IMDEA Nanociencia, C/Faraday 9, Campus de Cantoblanco, E-28049 Madrid, Spain.*

E-mail: andres.castellanos@imdea.org, emilio.perez@imdea.org

**Synthesis and characterization of perylenediimide (PDI) and tetraphenyl porphyrin (TPP)**

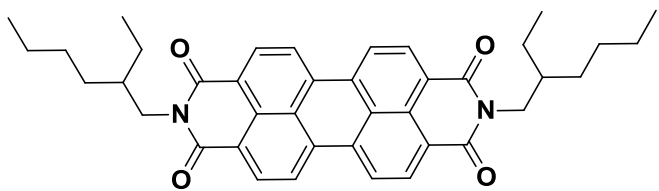

The **N,N'-bis(2-ethylhexan-1-amine)perylene-3,4,9,10-tetracarboxylic dianhydride** was synthesized and characterized as describe in *J. Org. Chem.* **2015**, *80*, 3036-3049, and showed identical spectroscopic data to those reported therein.

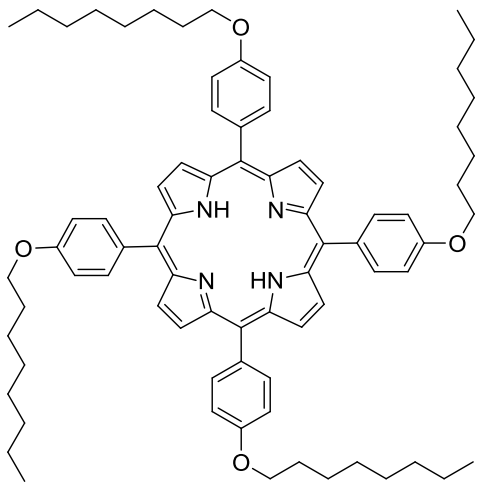

**5,10,15,20-tetra-(4-octyloxyphenyl)porphyrin.** 5,10,15,20-tetra-(4-hydroxyphenyl)porphyrin (500 mg, 0.74 mmol) was dissolved in dry DMF (100 mL) under argon and K$_2$CO$_3$ (2.65 g, 19.16 mmol) was added. 1-bromooctane (1.27 mL, 7.37 mmol) was added dropwise and the resulting mixture was stirred under reflux overnight. The mixture was poured onto cold HCl 1N and the solid was removed by filtration and dissolved in CHCl$_3$, then washed with water. The organic phase was dried over MgSO$_4$ and the solvent was removed under vacuum, obtaining the pure product. This compound (708 mg, 85% yield) was characterized by $^1$H, $^{13}$C-NMR, MALDI-TOF.

$^1$**H NMR** (400 MHz, CDCl$_3$) δ -2.68 (s, 2H), 0.97 (t, *J* = 6.9 Hz, 12H), 1.38 - 1.55 (m, 32H), 1.66 (q, *J* = 7.6 Hz, 8H), 2.01 (q, *J* = 6.8 Hz, 8H), 4.28 (t, *J* = 6.5 Hz, 8H), 7.30 (d, *J* = 8.8 Hz, 8H), 8.13 (d, *J* = 8.8 Hz, 8H), 8.89 (s, 8H).

$^{13}$**C NMR** (101 MHz, CDCl$_3$) δ 14.2, 22.8, 26.3, 29.4, 29.5, 31.9, 68.3, 112.7, 119.9, 134.5, 135.6, 158.9.

**MS** m/z: calculated for C$_{76}$H$_{94}$N$_4$O$_4$ 1127.6, found MALDI 1127.7.

**Field-effect characteristics of the functionalized MoS$_2$-based devices**

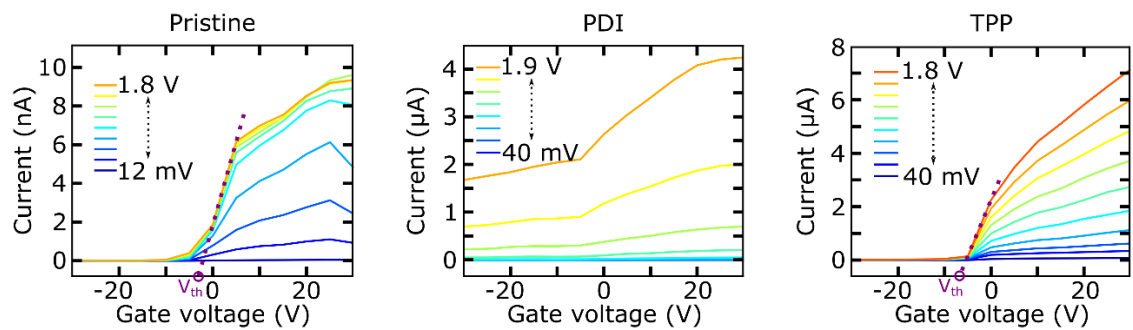

**Figure S1** Current-gate voltage traces for different drain-source voltages of the device shown in the main text measured in the pristine device, the PDI-coated device and the TPP-coated device.

The current-gate voltage traces are measured in air and in dark conditions. The ON/OFF ratio in the pristine device is 400, with a mobility of $3.4 \cdot 10^{-3}$ cm$^2$/V·s and a threshold voltage of ~ -3 V. When the device is coated with PDI, the ON/OFF ratio decreases approximately a factor of 100, the mobility increases a factor of 1000 and the threshold voltage shifts below -30 V, clearly showing a high n-doping due to the presence of the molecules. When the device is coated with TPP, the ON/OFF ratio remains the same, the mobility increases a factor of 1000 and the threshold voltage shifts to ~ -8 V, indicating a moderate n-type doping. Thus for TPP we attribute the current enhancement to be dominated by a reduction of the Schottky barrier height induced by the molecule/MoS$_2$ charge transfer rather than the direct n-type doping.

| Table S1. Field-effect characteristics of the MoS$_2$-based devices | | | |
|---|---|---|---|
|  | Mobility (cm$^2$/V·s) | $V_{th}$ (V) | ON/OFF ratio |
| Pristine | 3.4·10$^{-3}$ | -3 | 460 |
| PDI | 0.3 | < - 30 | 10 |
| Cleaned | 0.5·10$^{-3}$ | 0 | 30 |
| PDI | 1.11 | < - 30 | 10 |
| Cleaned | 0.3·10$^{-3}$ | -2 | 20 |
| PDI | - | - | - |
| Cleaned | - | - | - |
| TPP | - | - | - |
| Cleaned | 6.6·10$^{-3}$ | 0 | 1200 |
| TPP | 1.1 | -8 | 400 |
| Cleaned | - | - | - |
| TPP | 0.6 | -10 | 10 |
| Cleaned | 7.4·10$^{-3}$ | 0 | 2200 |

**Table S1** Field effect characteristics of the device shown in the main text at the different cycles of coating-cleaning-functionalization.

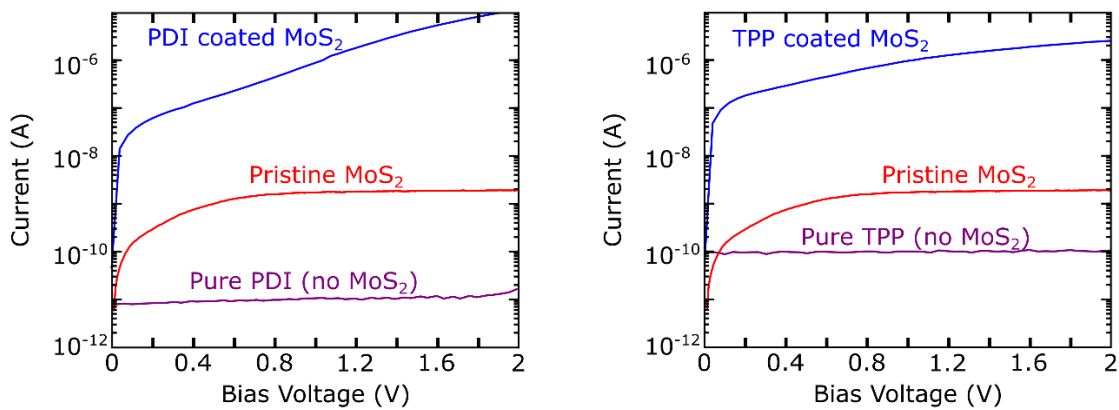

**Figure S2** Comparison of the current-voltage curves measured in the pristine device, the coated device and the molecules without MoS$_2$ bridging the electrodes. It can be seen that the molecules are highly resistive when compared to the coated device.

**UV-VIS spectra of PDI and TPP**

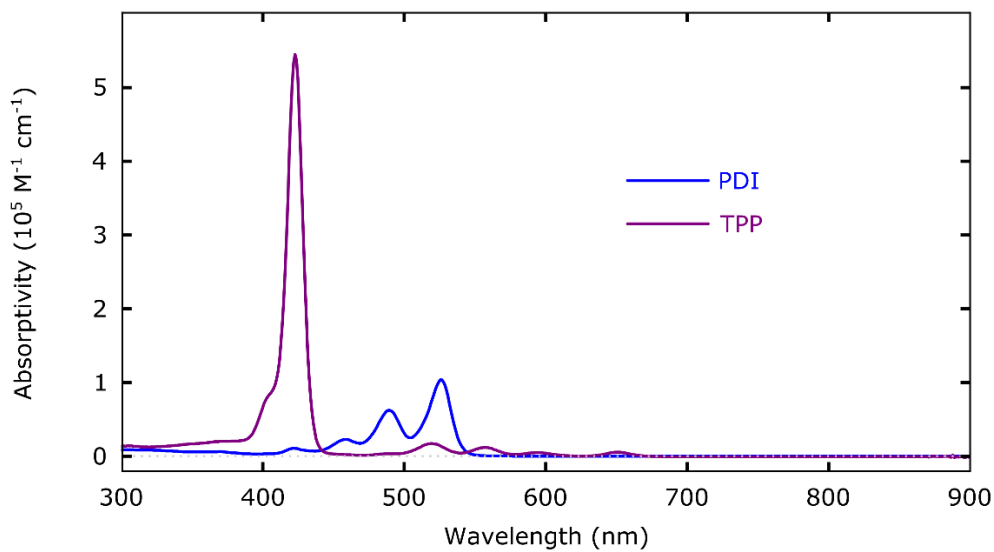

**Figure S3** UV-VIS absorptivity spectra of PDI and TPP molecules in CH$_2$Cl$_2$ solution.

**Optoelectronic characteristics of the functionalized MoS₂-based devices**

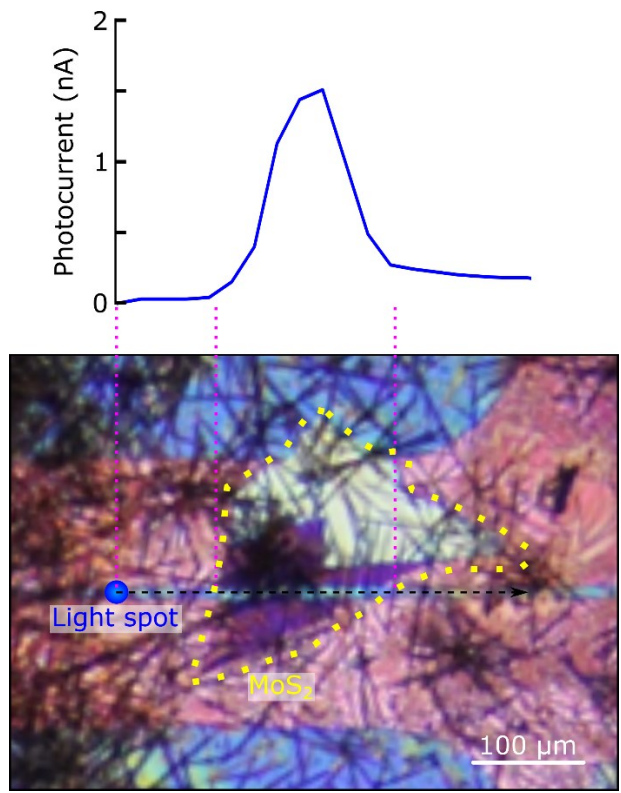

**Figure S4** Scanning photocurrent of the PDI-coated device with light wavelength of 455 nm. The light spot (diameter of 25 μm) is displaced over the sample while the current between the source and drain electrodes is measured. As it can be seen, when the light spot is outside the MoS₂ flake there is not photocurrent generation.

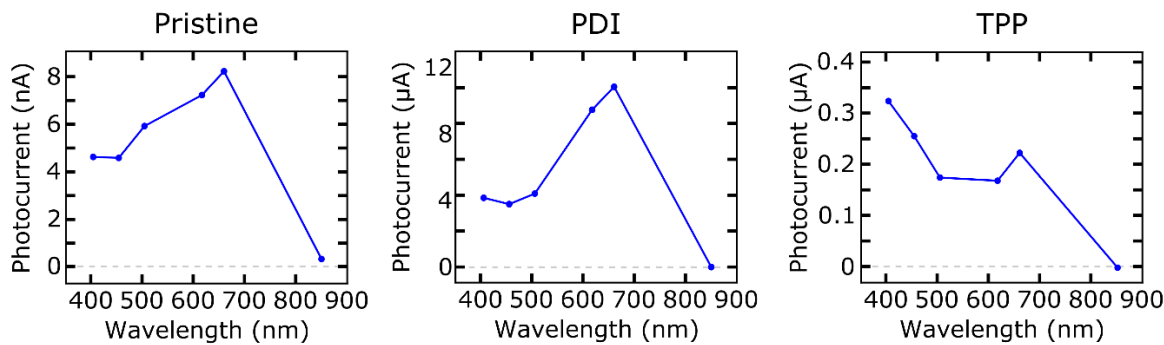

**Figure S5** Photocurrent measured in the device shown in the main text as a function of the light wavelength for the pristine device, the TPP-coated device and the PDI-coated device. The LED power is 100 nW and the photocurrent is measured with $V_{ds}$ = 2 V and $V_g$ = 30 V.

As it is shown in the plots, the device is responding to light for wavelengths shorter than 660 nm, where there is a photocurrent peak related to the MoS$_2$ A exciton. The photocurrent is enhanced in the coated device about 3 orders of magnitude with respect to the pristine device, although in the TPP-coated device the spectrum seems to be on top of a background which could be due to a high absorption of the TPP at high energies.

**Differential reflectance of functionalized MoS$_2$**

In Figure S6 we show optical microscopy and atomic force microscopy (AFM) topographic images of a monolayer MoS$_2$ flake transferred onto a glass substrate before and after functionalization. As can be seen in the AFM profile, the thickness of the pristine MoS$_2$ flake changes from 0.7 nm (monolayer) to 5-36 nm in the functionalized material due to the presence of a thin layer of TPPs.

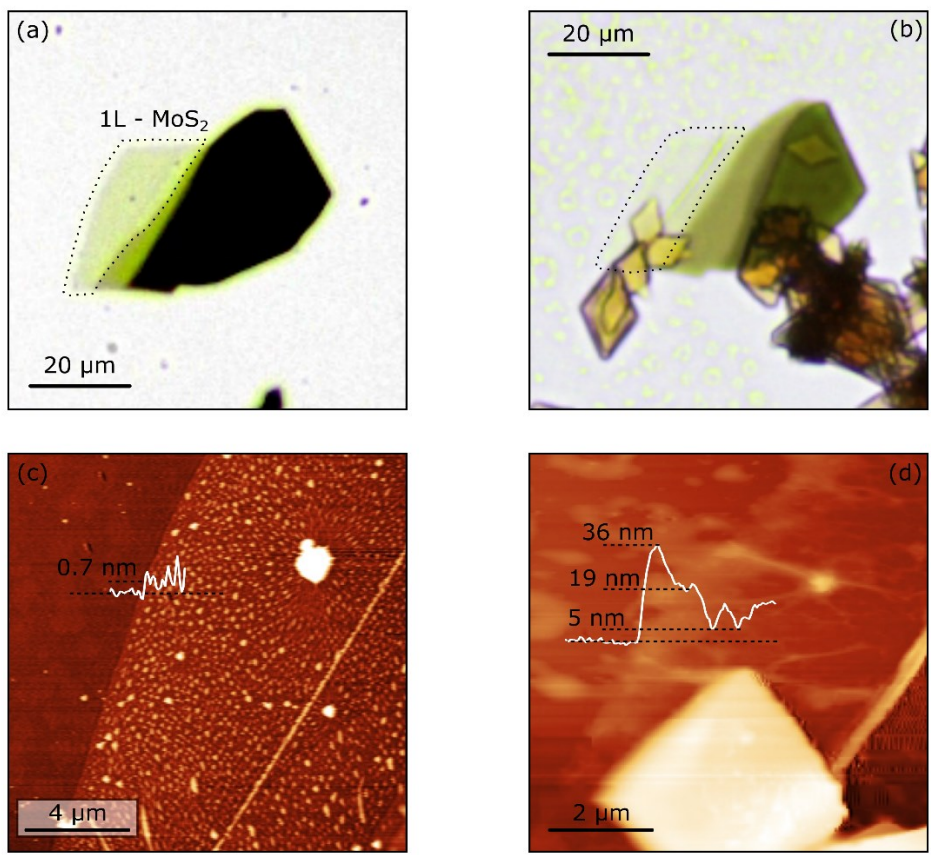

**Figure S6** Optical microscopy images of **(a)** the pristine MoS$_2$ flake and **(b)** after functionalization with TPPs. Atomic force microscopy topographic images of **(c)** the pristine MoS$_2$ flake and **(d)** after functionalization reveal that MoS$_2$ flake is covered by a thin layer (from 5 nm to 36 nm) of TPPs.

The differential reflectance spectrum, which measures the difference in reflectance of the MoS$_2$ flake and the glass substrate and is related to the absorption of the material, is shown in Figure S7a for the pristine and the functionalized MoS$_2$. In this spectra, the two peaks at 1.89 eV ± 0.01 eV, 2.03 ± 0.01 eV are due to the generation of the A and B excitons, associated to the optical transitions at the K point of the Brillouin zone. The only appreciable change is the addition of a rather featureless background of 5-10% in differential reflectance in the TPP functionalized sample.

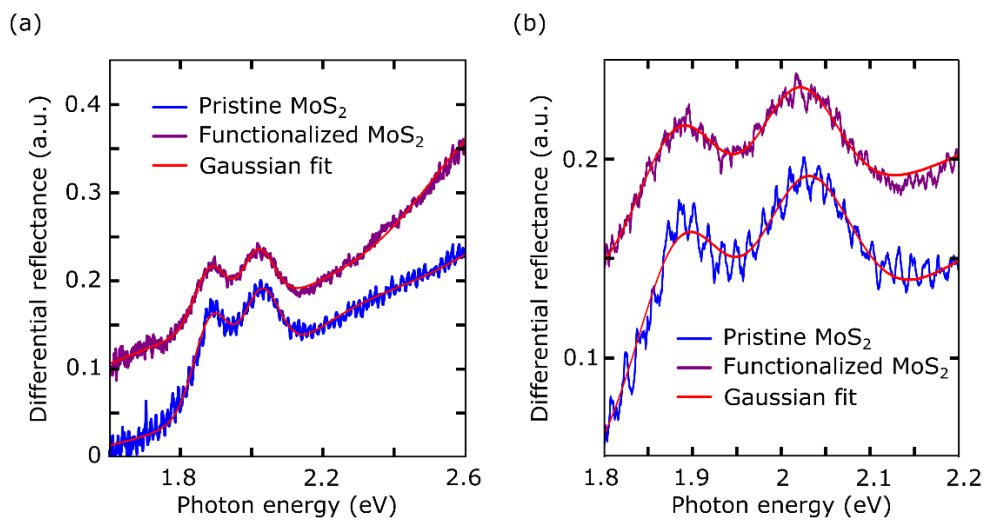

**Figure S7 (a)** Differential reflectance spectra of the pristine MoS$_2$ flake shown in Figure S6 and the functionalized flake with TPP molecules. **(b)** Gaussian fit of the main peaks appearing in (a). The fitted curves are centered at 1.89 eV ± 0.01 eV (A exciton) and 2.03 ± 0.01 eV (B exciton) in the pristine device and at 1.89 eV ± 0.01 eV and 2.02 ± 0.01 eV, respectively.